\documentclass[final]{siamltex}

\usepackage[dvips]{graphicx}
\usepackage{amsmath}
\usepackage{threeparttable}
\usepackage{multirow}
\usepackage{amsmath,amsfonts,amssymb,amscd,bm}

\title{A Fast Tree Algorithm for Electric Field Calculation in Electrical Discharge Simulations}

\author{Chijie Zhuang\thanks{State Key Lab of Power Systems and Department of Electrical Engineering, Tsinghua University, Beijing 100084, China. Supported by the national science foundation of China under grant 51577098. ({\tt chijie@tsinghua.edu.cn})}
\and Yong Zhang\thanks{Courant Institute of Computational Science, New York University. Supported by the Schr\"{o}dinger Fellowship J3784-N32.({\tt sunny5zhang@gmail.com})}
 \and Xin Zhou\thanks{Department of Electrical Engineering, Tsinghua University, Beijing 100084, China. }
 \and Rong Zeng\thanks{State Key Lab of Power Systems and Department of Electrical Engineering, Tsinghua University, Beijing 100084, China.  Supported by the national science foundation of China under grant 51325703 and 51377094.({\tt zengrong@tsinghua.edu.cn})}
 \and Jinliang He\thanks{Department of Electrical Engineering, Tsinghua University, Beijing 100084, China. ({\tt hejl@tsinghua.edu.cn})}
   \and Lei Liu\thanks{China Southern Grid Electric Power Research Institute, Guangzhou, 510663, China.}
}

\begin{document}

\maketitle

\begin{abstract}
The simulation of electrical discharges has been attracting a great deal of attention.
In such simulations, the electric field computation dominates the computational time.
In this paper, we propose a fast tree algorithm that helps to reduce the time complexity from
$O(N^2)$ (from using direct summation) to $O(N\log N)$. The implementation details are discussed and the time complexity is analyzed. A rigorous error estimation shows the error of the tree algorithm decays exponentially with the number of truncation terms and can be controlled adaptively. Numerical examples are presented to validate the accuracy and efficiency of the algorithm.
\end{abstract}

\begin{keywords}
tree algorithm, electric field, electrical discharge, disc model, error estimation
\end{keywords}


\pagestyle{myheadings}
\thispagestyle{plain}
\markboth{CHIJIE ZHUANG AND YONG ZHANG et al.}{A Fast Tree Algorithm for Electric Field Calculation in Electrical Discharge Simulations}

\section{Introduction}
There are various types of electrical discharges in nature, e.g., lightning strikes \cite{iet}, corona discharges around electrodes in non-uniform electric fields \cite{onset,trichel}. Because of the relevance of electrical discharge to everyday life and its growing application in industry, the numerical simulation of electrical discharges has been increasingly attracting attention.

The most widely adopted model for electrical discharge simulations is the fluid model \cite{dg,jcp}. This model consists of the Poisson equation, which describes the electric field that drives the electrical discharge, and the convection-diffusion equations with source terms, which describe the charge-carrier transport.

Because of its high computational load, the simulation of electrical discharge under atmospheric pressure is, at present, mainly limited to short gap discharges of a few centimeters in length \cite{jcp}. Thus, many simplified models have been proposed in the hope of simulating longer discharges, e.g., 100 cm in length. Among these models, the most promising one is the so-called 1.5-dimensional model \cite{amc}.

\begin{figure}[!htp]
  \centering
  \includegraphics[width=3.5in]{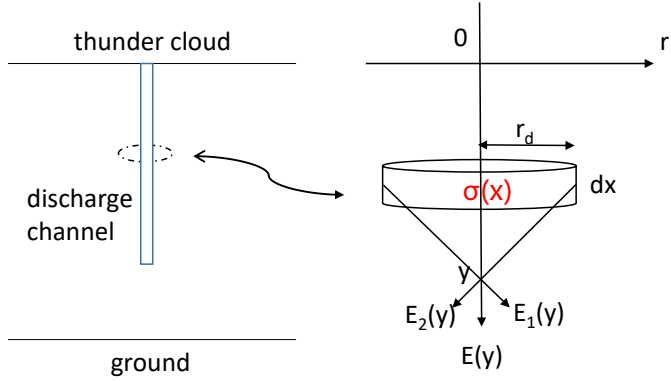}
  \caption{Diagram of a 1.5-dimensional model.}\label{15dm}
\end{figure}

In the 1.5-dimensional model, the charges are assumed to be distributed among discs of the same radius. On each disc the charge density is uniform, and the charges only move along the $y$-axis (see Fig. \ref{15dm}). The charge transport is described using a one-dimensional model, while the electric field is considered to be two-dimensional. Using this so-called disc method, the electric field can be derived analytically. Assume there is a disc of net charge density $\sigma(x)$, radius $r_d$, thickness $\mbox{d}x$ (see Fig. \ref{15dm}), and the permittivity of air is $\varepsilon_0$. The electric field it generates at a point, $y$, along the $y$-axis is given by \cite{amc}:
\begin{equation}
\mbox{d}E(y)=\Bigg\{
\begin{array}{l l}
\frac{\sigma(x)}{2\varepsilon_0}(\frac{x-y}{\sqrt{(y-x)^2+r_d^2}}+1)\mbox{d}x,& x-y < 0;\\
\frac{\sigma(x)}{2\varepsilon_0}(
\frac{x-y}{\sqrt{(y-x)^2+r_d^2}}-1)\mbox{d}x,& x-y\ge 0.
\end{array}
\end{equation}

{To consider the influence of the electrodes on the electric field, all image charges, e.g., which are above the cloud and below the ground in Fig. \ref{15dm},}  should be taken into account. However, only image charges whose distances to the electrodes are less than the discharge-gap length $L$ are considered because image charges that are far way contribute little to the electric field. Integrating over the whole domain, we get
\begin{eqnarray} \label{eq2}
E(y)&=&\frac{1}{2\varepsilon_0}\bigg[\int_{-L}^{y} \sigma
(x)\bigg(\frac{x-y}{\sqrt{(x-y)^2+r_d^2}}+1\bigg)\mbox{d}x \nonumber \\
&&+\int_{y}^{L}\sigma(x)\bigg(\frac{x-y}{\sqrt{(x-y)^2+r_d^2}}-1\bigg)\mbox{d}x\bigg].
\end{eqnarray}

Assuming there are $N$ source charges and $N$ target points, the computation of Eq. (\ref{eq2}) has a time complexity of $O(N^2)$. As a result, the electric field evaluation may occupy around 90\% of the CPU time in a simulation \cite{pv}, and fast algorithms with better complexity are highly imperative.

{
In fact, there has been many works on such fast evaluation of potential and field, e.g., the
Barnes--Hut fast tree algorithm \cite{tree1}, and the famous fast multipole method \cite{fmm,fmmshort} for
$N$-body simulation. In this paper, we propose a tree algorithm for the specific kernel arising from electrical discharge simulations, employing the same ideas of far-field, near-field evaluation, which dramatically helps to accelerate the field evaluation with a highly controllable accuracy \cite{tree1,tree2}.
}

\section{The Tree Algorithm}
By integrating Eq. (\ref{eq2}) using sufficient high-order Gaussian quadrature, and setting $q_j := \frac{\omega_j \sigma_j}{2\varepsilon_0} \Delta x$ where $\omega_j$ is the associated weight of the Gaussian quadrature and $\Delta x$ is the length of the associated interval, Eq. (\ref{eq2}) can be reduced to
\begin{eqnarray}
   E(y) &=& \big(\sum_{j=0}^m q_j-\sum_{j=m+1}^n q_j\big)+\sum_{j=0}^n \frac{q_j(x_j-y)}{\sqrt{(x_j-y)^2+r_d^2}} \nonumber \\
   &:=&e_{m}+\sum_{j=0}^n \frac{q_j(x_j-y)}{\sqrt{(x_j-y)^2+r_d^2}}.
\end{eqnarray}
where $e_{m}=\sum_{j=0}^m q_j-\sum_{j=m+1}^n q_j$.
The term $e_m$ can be calculated recursively, i.e.,
\begin{equation}
  e_{m+1}= e_m+2q_{m+1}. \nonumber
\end{equation}
Thus $e_0$ is computed first, followed by the successive calculation of $e_1$, $e_2$..., $e_n$. This work has a linear time complexity.
Below we will omit the term $e_m$ for brevity, but the principle of the tree algorithm remains unchanged.

\begin{figure}[!htp]
  \centering
  \includegraphics[width=2.5in]{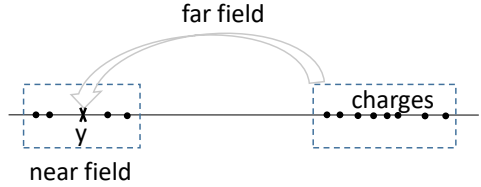}
  \caption{Diagram of the near-field and far-field interactions.}\label{treediag}
\end{figure}

As shown in Fig. \ref{treediag}, the total electric field, $E$, is split into two parts, i.e., the far-field $E_f$ and the near-field $E_n$ such that $E=E_f+E_n$. The fundamental idea of the tree algorithm is that the far-field interaction, which is from the charges far away from the target point, is approximated as if they are a group, while the near-field from the neighboring charges is evaluated directly.

Assume a cluster of charges $\{q_j\}_{j=0}^n$ located at $\{x_j\}_{j=0}^n$ are gathered around $x_c$, and $|y-x_c|\gg 0$, $|y-r_d|\gg 0$. To calculate the far-field $E_f(y)$,
a crude approximation is
\begin{eqnarray}
  E_f(y) = \sum_{j=0}^n q_j \Phi(x_j,y) \approx \big(\sum_{j=0}^n q_j\big)\Phi(x_c,y),\label{eqn}
\end{eqnarray}
with $\Phi(x,y) := \frac{x-y}{\sqrt{(x-y)^2+r_d^2}}$. However, using Taylor expansion, we have
\begin{eqnarray}
  \Phi(x,y) &=&\sum_{k=0}^{\infty} \frac{1}{k!} \Phi^{(k)}(x_c, y) (x-x_c)^k \nonumber\\
  &=&\sum_{k=0}^p \frac{1}{k!} \Phi^{(k)}(x_c, y) (x-x_c)^k + R_p(x), \label{eqtaylor}
\end{eqnarray}
where $\Phi^{(k)} = \frac{\partial^{k} \Phi}{\partial x^k}$; $p\in \mathbb{N}$; the residual $R_p$ is given by $R_p=\sum_{k=p+1}^\infty \frac{1}{k!} \Phi^{(k)}(x_c, y) (x-x_c)^k$. Therefore, we have
\begin{eqnarray}
  E_f(y)
  &=& \sum_{j=0}^n q_j \left(\sum_{k=0}^{\infty} \frac{1}{k!} \Phi^{(k)}(x_c, y) (x_j-x_c)^k\right) \nonumber \\
  &\approx& \sum_{k=0}^{p} \Phi^{(k)}(x_c, y)\left(\sum_{j=0}^n q_j \frac{(x_j-x_c)^k}{k!}\right). \label{eqa}
\end{eqnarray}
When $p=0$, Eq. (\ref{eqa}) reduces to the crude approximation Eq. (\ref{eqn}).
To approximately calculate $E_f(y)$, one only needs to calculate the moments $\left(\sum_{j=0}^n q_j \frac{(x_j-x_c)^k}{k!}\right)$ and $\Phi^{(k)}(x_c, y)$, for $k=0,...p$.

We now derive a recurrence formula to calculate $\Phi^{(k)}(x,y)$. It is straightforward that
\begin{eqnarray}
 &\Phi^{(0)}(x,y)= \frac{x-y}{\sqrt{(x-y)^2+r_d^2}} \label{eqm1}, \\
 &\Phi^{(1)}(x,y)= \frac{r_d^2}{\big(\sqrt{(x-y)^2+r_d^2}\big)^3},\label{eqm2}
\end{eqnarray}
which implies that
\begin{equation}
   r_d^2 \Phi^{(0)}(x,y)= \Phi^{(1)}(x,y) {\big[}(x-y)^3+r_d^2 (x-y){\big ]} \label{eqphi}.
\end{equation}
Differentiating Eq. (\ref{eqphi}) for $k$-1 times using the general Leibniz rule, after some algebraic simplifications, we get
\begin{eqnarray}
(x-y){[}(x-y)^2+r_d^2{]} \Phi^{(k)}(x,y)=\nonumber \\
{\big[}r_d^2 -(k-1)(3(x-y)^2+r_d^2){\big]}\Phi^{(k-1)}(x,y)\nonumber \\
-3(k-1)(k-2)(x-y)\Phi^{(k-2)}(x,y)\nonumber \\
-(k-1)(k-2)(k-3)\Phi^{(k-3)}(x,y). \label{eqm3}
\end{eqnarray}

Therefore, by using Eqs. (\ref{eqm1}) to (\ref{eqm3}), for any given $y$, $\Phi^{(k)}(x_c, y)$ may be calculated recursively for $k=2,3,...p$.

\section{Error Estimation}
Now we present a rigorous error estimation for Eq. (\ref{eqa}).
Without loss of generality, we only consider the case $x_c=0$.
Other cases reduce to the $x_c=0$ case after a simple shift, i.e. let $x:=x-x_c$.

Define a complex function $f(z):=\frac{z-y}{\sqrt{(z-y)^2+r_d^2}}$ with $z \in \mathbb{C}$,
which is analytic for $|z|< \sqrt{y^2+r_d^2}$.
By Cauchy's integral formula, for any $z$ satisfying $|z|:=r\leq R:=|y|$,
\begin{eqnarray}
  f(z)=\frac{1}{2\pi i} \oint_\Gamma \frac{f(\xi)}{\xi-z}\mbox{d}\xi,~~~~
  f^{(k)}(0) = \frac{k!}{2\pi i} \oint_\Gamma\frac{f(\xi)}{\xi^{k+1}}\mbox{d}\xi,
\end{eqnarray}
where $i=\sqrt{-1}$, $\Gamma:=\{w \in \mathbb{C}||w|=R\}$ is a contour containing the point $z$.
We have
\begin{eqnarray}
 f(z)&=& \frac{1}{2\pi i} \oint_\Gamma \frac{f(\xi)}{\xi}\frac{1}{1-\frac{z}{\xi}}\mbox{d}\xi \nonumber \\
  &=& \frac{1}{2\pi i} \oint_\Gamma\frac{f(\xi)}{\xi}\big(\sum_{k=0}^p (\frac{z}{\xi})^k+\frac{(\frac{z}{\xi}) ^{p+1}}{1-\frac{z}{\xi}}\big)\mbox{d}\xi \nonumber \\
  &=& \frac{1}{2\pi i} \big(\sum_{k=0}^p z^k\oint_\Gamma \frac{f(\xi)}{\xi^{k+1}}\mbox{d}\xi+\oint_\Gamma \frac{f(\xi)}{\xi}\frac{(\frac{z}{\xi}) ^{p+1}}{1-\frac{z}{\xi}}\mbox{d}\xi\big) \nonumber \\
  &=& \sum_{k=0}^p \frac{f^{(k)}(0)}{k!} z^k +\frac{1}{2\pi i}\oint_\Gamma \frac{f(\xi)}{\xi}\frac{(\frac{z}{\xi}) ^{p+1}}{1-\frac{z}{\xi}}\mbox{d}\xi. \label{eqcauchy}
\end{eqnarray}
Comparing $f(z)$ and $\Phi(x)$, we find that $\Phi(x) = f(z)|_{z=x}$, so
\begin{eqnarray}
  |R_p|&=&\left|\frac{1}{2\pi i}\oint_\Gamma \frac{f(\xi)}{\xi}\frac{(\frac{z}{\xi}) ^{p+1}}{1-\frac{z}{\xi}}\mbox{d}\xi\right|_{z=x} \nonumber \\
  &\leq& \max\left\{\left|\frac{1}{2\pi i}\oint_\Gamma \frac{f(\xi)}{\xi}\frac{(\frac{z}{\xi}) ^{p+1}}{1-\frac{z}{\xi}}\mbox{d}\xi\right|\right\}.
\end{eqnarray}

Using the fact $|f(\xi)|$ is bounded for $\xi \in \Gamma$, i.e. $\left|f(\xi)\right|\leq M$, we get
\begin{eqnarray}
  |R_p|
  &\leq& \frac{1}{2\pi}\oint_\Gamma \max\big(\left|\frac{(\frac{z}{\xi})^{p+1}}{1-\frac{z}{\xi}}\right|\left|\frac{f(\xi)}{\xi}\right|\big)\mbox{d}\xi \nonumber\\
  &\leq& \max{\big(\left|\frac{(\frac{z}{\xi})^{p+1}}{1-\frac{z}{\xi}}\right|\big)} \max\left| f(\xi)\right| \nonumber\\
  &\leq& M\frac{R}{R-r} (\frac{r}{R})^{p+1}. \label{error}
\end{eqnarray}

Eq. (\ref{error}) shows Eq. (\ref{eqa}) converges as $p$ increases if $r<R$, which is easy to be satisfied; to be more precise, the error decays exponentially with respect to $p$. As an example, $|R_p|$ is sufficiently small when $p=15$ or 20 if $\frac{r}{R}\leq\frac{1}{3}$.

\section{Implementation and Efficiency Analysis}
\begin{figure}[!htp]
  \centering
  \includegraphics[width= 3.0in]{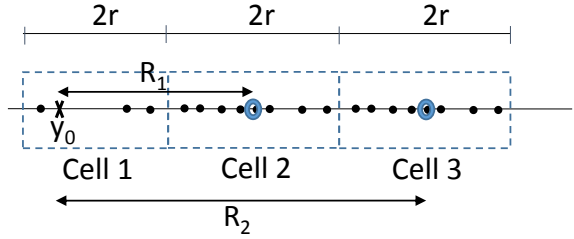}
  \caption{Diagram showing direct and well-separated neighbors.}\label{far-near}
\end{figure}

Equation (\ref{eqa}) is used to approximate the far field when the target and sources points are well separated.
Now we illustrate in Fig.~\ref{far-near} how to determine whether the target and source charges are well separated.
In Fig.~\ref{far-near}, three intervals, all with a diameter of $2r$, are shown.
The target point, $y_0$, lies in cell 1, and its distances to centres of cell 2 and 3 are $R_1$ and $R_2$ respectively.
We say that cell 1 and cell 2 are direct neighbors if $\frac{r}{R_1}>\frac{1}{3}$; while cell 1 and cell 3 are well separated if and only if $\frac{r}{R_2}\leq\frac{1}{3}$.

\begin{figure}[!htp]
  \centering
  \includegraphics[width= 3.25in]{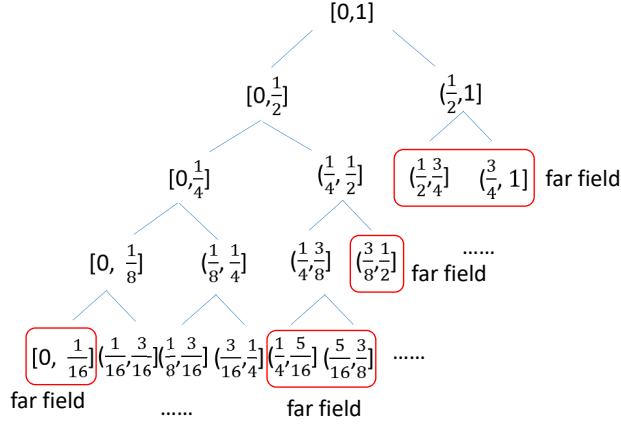}
  \caption{Diagram of the binary tree structure.}\label{treestructure}
\end{figure}

We now build a binary tree to successively approximate the far field, an example of which can be seen in Fig. \ref{treestructure}. For simplicity, all the sources are assumed to be in [0,1] and the target is assumed to be in $[\frac{1}{8}, \frac{3}{16}]$. The target point and all the charges are direct neighbors at the first two levels. In level 2, the target and $(\frac{1}{2}, \frac{3}{4}]$, $(\frac{3}{4}, 1]$ are well separated while all others remain direct neighbors. The intervals are further subdivided, which results in $(\frac{3}{8}, \frac{1}{2}]$ becoming the well separated neighbor.

This process is repeated until the bottom level is reached. There are finally at most two direct neighbors of the interval containing the target point, while the very interval and all other intervals are well separated. The near field from the interval containing the target point and the direct neighbors, is evaluated directly; while the far field from other well separated intervals at different levels are approximated by Eq. (\ref{eqa}).

Assuming $c$ is the number of source charges in an interval at the bottom level of the tree, then $c$ is $O(1)$ and $2^m\approx N/c$, which implies $m$  is $O(\log N)$.

Now we are ready to estimate the computational cost for a single target point, ignoring the cost of setup.
The work for the far field part involves the evaluation of at most three far field expansions of $p$ terms at each
level from 2 to $m$. Therefore, the flop count arising from the evaluation of the far
field expansions is $O(mp)$. The near field evaluation, which is done at the bottom level, requires at most three intervals. Since each bottom level interval contains only $O(1)$ sources, the flop count of the direct calculation is $O(1)$.
Hence the cost in flops of an evaluation for one single target point is $O(mp)\approx O(p\log N)$,
which is typically much faster than the $O(N)$ flop count associated with
direct summation.

At each level, the setup cost, which is mainly the formation of the moments in Eq. (\ref{eqa}), is at most $O(pN)$, so the total setup cost is at most $O(mN)\approx O(N\log N)$.

Overall, the computational cost of the algorithm is $O(N\log N)$ for $N$ targets.

\section{Validation and Efficiency}
\subsection{{Variation of the error with the number of truncation terms}}
First, only the far-field was calculated to validate Eq. (\ref{eqa}).
We randomly generated 10000 charges in $[-0.5,0.5]$, which is around $x=0$, and set $r_d=0.1$, The electric field at $y=1$ ($r/R=0.5$) was then calculated using different numbers of truncation terms, denoted by $p$.
Results in Tab. \ref{tab1} show the relative error decays exponentially with $p$, which coincides with the error estimation in Eq. (\ref{error}).
When $p$ increases by five, the error decays by a factor of about 50-100.

{
\begin{table}[!h]
\caption{Accuracy with different numbers of truncation terms ($p$)}
\label{tab1}
\centering
\begin{tabular}{c|cccc}
\hline
{$p$}  & 5 & 10  & 15 &20 \\
\hline
{relative error} & 9.43e-5 & 1.17e-6 & 6.40e-8 &6.48e-10 \\
\hline
\end{tabular}
\end{table}
}

\subsection{{Impact of the number of tree levels on efficiency}}
Next, the impact of the number of tree levels on the CPU calculation time was tested. The algorithm was implemented in C++, {and the experiments here and below were performed on a PC with an i7-6500U CPU and 8 GB RAM.} With more levels, the effect of more sources are calculated by Eq. (\ref{eqa}), which may accelerate the computation; however, more tree levels are traversed, which may slow down the computation. In our experiment, $2\times 10^5$ charge sources, each with a random amount of charge, were uniformly randomly distributed in $[0,1]$, the target and source positions were the same. The field generated by the neighboring charges were calculated directly and others by Eq. (\ref{eqa}) with $p=10$.

It is shown in Tab. \ref{tab2} that the depth of the tree or, in other words, the number of particles in the bottom-level interval, greatly influences the computational efficiency. Our test shows that the best number of particles is about 48 for $p=10$.

{
\begin{table}[!h]
\caption{Computational time with different depths($p=10$, $N=2\times10^5$)}
\label{tab2}
\centering
\begin{tabular}{c|ccccccccc}
\hline
{levels} &9  &10 &11 &12 &13 &14 &15 &16\\
\hline
{\# particles } & 781 & 390 & 195 & 97  & 48  & 24  & 12  & 6  \\
{time (ms)} &  2418  &  1450  &  1014 &  827   &  765   &  780   &  795   &  842  \\
\hline
\end{tabular}
\begin{tablenotes}
        \footnotesize
        \item[1] \# particles means the estimated number of the particles in one interval at the bottom level of the tree.
\end{tablenotes}
\end{table}
}

\subsection{Efficiency with different number of particles and targets}
After optimizing the number of particles in the bottom-level interval, both the near-field and far-field were evaluated in order to test the efficiency of the tree algorithm. Different numbers of charge sources, each with a random amount of charge, were randomly placed in $[0,1]$, the evaluation locations were the same as the source positions. The tree levels were determined such that the finest interval contained about 40 particles. The other configuration were the same as in the above experiment and the experiments were repeated multiple times. The results in Tab. \ref{tab3} show that the time complexity of the algorithm is roughly $O(N\log N)$, which is much faster than direct summation even when $N$ is small.

\begin{table}[!h]
\caption{Time-cost comparisons for different numbers of charges and targets}
\label{tab3}
\centering
\begin{tabular}{c|c|c|c|c|c|c}
\hline
\multirow{2}{*}{particles} & \multicolumn{3}{c|}{time by tree algorithm (ms)} & \multicolumn{3}{c}{ time by direct summation (ms)} \\
\cline{2-7}   & max & min & average & max & min & average \\
\hline
1e4 & 29 & 24 & 27.4 & 577 & 453 & 505.5   \\
5e4 & 170& 156& 160.4 & 11840 & 11060 & 11372.8 \\
10e4& 385 & 353 & 362.8 & 48518 & 44058 & 45929.4 \\
15e4& 587 & 571 & 577.4 & 104083 & 100843 & 101882.8 \\
20e4& 858 & 674 & 772.8 & 188105 & 180105 & 182944.8\\
\hline
\end{tabular}
\end{table}

In addition, two types of errors, maximal (left) and average (right), were measured:
\begin{eqnarray}
 \max_i\left|\frac{E^{\mbox{tree}}_i-E^{\mbox{dir}}_i}{E^{\mbox{dir}}_i}\right|,\quad
 \quad \frac{\sum_i(|E^{\mbox{tree}}_i-E^{\mbox{dir}}_i|)}{\sum_i |E^{\mbox{dir}}_i|}. \nonumber
\end{eqnarray}

Table \ref{tab4} shows the maximal and average errors are roughly of the same order for different numbers of particles and targets, and are very small, which infers that the algorithm is reliable.

\begin{table}[!h]
\caption{Accuracy for different numbers of charges and targets}
\label{tab4}
\centering
\begin{tabular}{l|ccccc}
\hline
$\#$ of particles &1e4&5e4 &10e4 &15e4 &20e4\\[0.25em]
maximal error & 2.89e-10  & 4.41e-10   & 2.44e-9   & 1.61e-9   & 2.75e-9  \\[0.2em]
 average error& 6.16e-14   & 4.96e-14 & 4.90e-14  & 4.75e-14 & 5.25e-14\\ \hline
\end{tabular}
\begin{tablenotes}
        \footnotesize
        \item[1] \# of particles means the total number of particles.
\end{tablenotes}
\end{table}

\section{Conclusion}

We present in this paper a fast tree algorithm of $O(N\log N)$ complexity to calculate the electric field arising from 1.5 dimensional electrical discharge simulations.

{
The tree algorithm is derived based on Taylor expansion. A recurrence formula following the general Leibniz rule is provided to calculate the expansion coefficients efficiently.

Error estimation shows that the error decays exponentially as the number of truncation terms increases and detailed analysis confirms the $O(N\log N)$ time complexity, which represents a dramatic improvement over direct summation method with tunable accuracy. Numerical experiments were given to validate the efficiency and accuracy.

Developing algorithms of linear time complexity following the ideas of fast multipole method will be our possible further direction, especially in higher space dimensions.
}

\section*{Acknowledgement}
This work is supported by the National Natural Science Foundation of China under grant 51577098, 51325703 and 51377094, Open Fund of National Engineering Laboratory for Ultra High Voltage Engineering Technology (Kunming, Guangzhou), the Schr\"{o}dinger Fellowship J3784-N32.
Prof. Jingfang Huang at University of North Carolina, Chapel Hill, is greatly acknowledged for continuous help during the years.

\end{document}